\documentclass[aps,twocolumn,showpacs,preprintnumbers,amsmath,amssymb,pre]{revtex4-1}
\usepackage{epsf,amsmath,amssymb,amsfonts,verbatim,color,multirow,pifont}
\usepackage{graphicx}

\begin{document}
\title{Additivity of multiple heat reservoirs in Langevin equation}
\author{Jae Sung Lee} \email{jslee@kias.re.kr}
\author{Hyunggyu~Park} \email{hgpark@kias.re.kr}
\affiliation{School of Physics and Quantum Universe Center, Korea Institute for
Advanced Study, Seoul 02455, Korea}

\newcommand{\revise}[1]{{\color{red}#1}}

\date{\today}

\begin{abstract}
 The Langevin equation greatly simplifies the mathematical expression of the effects of thermal noise by using only two terms, a dissipation term, and a random-noise term. The Langevin description was originally applied to a system in contact with a single heat reservoir; however, many recent studies have also adopted a Langevin description for systems connected to multiple heat reservoirs. This is accomplished through the introduction of a simple summation for the dissipation and random-noise terms associated with each reservoir. However, the validity of this simple addition has been the focus of only limited discussion and has raised several criticisms. Moreover, this additive description has never been either experimentally or numerically verified, rendering its validity is still an open question. Here, we perform molecular dynamics simulations for a Brownian system in simultaneous contact with multiple heat reservoirs to check the validity of this additive approach. Our simulation results confirm that the effect of multiple heat reservoirs is additive in general.
 A very small deviation in the total amount of dissipation and associated noise is found, but seems not significant within statistical errors. We find that the steady-state properties satisfy the additivity perfectly and are not affected by this deviation.
\end{abstract}


\maketitle

\section{Introduction}
\label{sec:intro}

The Langevin equation is a stochastic differential equation that describes the motion of a system as it interacts with a thermal reservoir. The exact mathematical expression for this system-reservoir interaction is often very complicated, making it difficult to define thermodynamic quantities in the strong interaction regime~\cite{Seifert2016,Jarzynski2017}. However, in the Langevin description, the effect of a thermal reservoir on a system is phenomenologically expressed through only two terms, a dissipation term and an associated random-noise term~\cite{Risken}. This simplification enables analysis of systems affected by thermal noise at the mesoscopic scale. As heat can be naturally, and perhaps intuitively, defined as work done by these two terms~\cite{Sekimoto,Noh,LeeJS}, we also see why the ratio of the logarithms of the forward and time-reversal path probabilities can be interpreted as entropy production, which is one of the core discoveries of recent developments in the field of stochastic thermodynamics~\cite{Seifert2005,Hatano,Spinney,Tome,Sagawa,Lee}. Additionally, the time evolution of the probability distribution function of such a system can be investigated through use of the corresponding Fokker–Planck equation.

Originally, the Langevin equation was introduced to describe the motion of a \emph{Brownian} particle with a single degree of freedom immersed in a \emph{single} heat reservoir (BS). Recently, there have been various studies investigating a \emph{Brownian} system with a single degree of freedom simultaneously connected to \emph{multiple} heat reservoirs (BM) but still using the Langevin equation~\cite{Derrida,Visco,Chun,Broeck1,Murashita,Parrondo}. These studies have been motivated by the development of Brownian motors~\cite{Broeck,Meurs,Kawai,Reimann,Hanggi} and the Feynman-Smoluchowski ratchet~\cite{Smoluchowski,Feynman,LeeJS1}. We explicitly distinguish BM from the multiple-heat-reservoir systems where a single degree of freedom is affected by only one heat reservoir at a time~\cite{Sinitsyn,Sourabh}.

Our examination of BM using the Langevin equation proceeds as follows. Suppose we have $n$ heat reservoirs and a Brownian particle. When the particle is connected only to reservoir $i$ ($i=1,\cdots,n$) with temperature $T_i$, its one-dimensional motion can be described as
\begin{eqnarray}
\dot{x}=v,~~m\dot{v} = f(x) -\gamma_i v + \xi_i, \label{eq:Lan_eq1}
\end{eqnarray}
where $x$, $v$, and $m$ are the position, velocity, and mass of the particle, respectively; $f(x)$ is an external force; and $\gamma_i$ is the dissipation coefficient associated with the $i$-th reservoir. $\xi_i$ is the Gaussian white noise term of the $i$-th reservoir, with statistical properties satisfying $\langle \xi_i(t) \xi_i(t^\prime) \rangle =2 D_i  \delta(t-t')$, where $t$ is time and $D_i=\gamma_i T_i$ is the noise strength of $\xi_i$. In the analysis, we set the Boltzmann constant $k_B =1$. We note that the expression $-\gamma_i v+ \xi_i$ describes the effect of the reservoir.

Now, we need to expand the system by imagining that the particle is in contact with all $n$ reservoirs simultaneously. What would a Langevin equation for such a system contain? Previous studies~\cite{Derrida,Visco,Chun,Broeck1,Murashita,Parrondo} have opted to treat the effect of the multiple reservoirs as additive processes, written as
\begin{equation}
\dot{x}=v,~~m\dot{v} = f(x) -\gamma_{1,\cdots,n} v + \xi_{1,\cdots,n}, \label{eq:Lan_eq2}
\end{equation}
where $\gamma_{1,\cdots,n} = \sum_{i=1}^n \gamma_i$ and $\xi_{1,\cdots,n} = \sum_{i=1}^n \xi_i$. Here, we note that $\langle \xi_{1,\cdots,n} (t) \xi_{1,\cdots,n} (t^\prime) \rangle =2 \delta(t-t') D_{1,\cdots,n}$, where $D_{1,\cdots,n}=\sum_{i=1}^n D_i $. Based on this equation, many thermodynamic problems of BM, such as the heat distribution~\cite{Visco}, the amount of irreversible heat flow~\cite{Parrondo}, the entropy production~\cite{Broeck1}, and the overdamped limit~\cite{Murashita}, have been studied.

However, the justification of the additivity of multiple heat reservoirs in this manner and, hence, the form of Eq.~\eqref{eq:Lan_eq2} is a non-trivial problem and remains an open question. This issue has been addressed theoretically in a highly specific situation~\cite{Broeck,Meurs}, where Eq.~\eqref{eq:Lan_eq2} was derived up to first order in the mass ratio between a reservoir particle and Brownian particle in the low (reservoir particle) density, or the large mean-free-path, regime. The additivity in more general situations has not yet been fully explored. There have also been criticisms on the assumption of additivity by H\"{a}nggi~\cite{Hanggi1}, who claimed that transient relaxation dynamics may not be sufficiently described by Eq.~\eqref{eq:Lan_eq2}, as initial condition dependence does not dampen away in transient dynamics. Furthermore, non-equilibrium dynamics with multiple reservoirs (which need not all be at the same temperature) should be distinguished from simple equilibrium dynamics at an effective temperature. Nevertheless, H\"{a}nggi agreed that the steady-state dynamical behavior will be correctly described by Eq.~\eqref{eq:Lan_eq2}, as the effect of the initial conditions become negligible over long times. Parrondo and Espa\~{n}ol~\cite{Parrondo} also stated that Eq.~\eqref{eq:Lan_eq2} is correct only for the asymptotic long-time limit, i.e., the steady state. To our knowledge, the additivity property has never been experimentally or numerically verified, making it important to check its validity in general situations.

To accomplish this task, we perform molecular dynamics (MD) simulations for BM (two reservoirs). We find that Eq.~\eqref{eq:Lan_eq2} describes the BM dynamics well in general. More specifically, there seems to exist a small raising in dissipation $\gamma_{1,2}$ and noise strength $D_{1,2}$ from the simple additivity. It should be noted that these small corrections appear even for the case of two separate reservoirs with the same temperature, thus cannot be attributed to non-equilibrium-ness with $T_1 \neq T_2$.
As their magnitudes are just comparable to or smaller than statistical errors,
quantitative investigation on its origin is not properly carried out in this study, which will be left for future study.
We report that the effective temperature $T_{1,2}=D_{1,2}/\gamma_{1,2}$, which characterizes the steady state, seems to be in agreement with that of Eq.~\eqref{eq:Lan_eq2} without any detectable deviation.

The remainder of this paper is organized as follows. We describe our model for the performed MD simulations in Sec.~\ref{sec:model}. The results of these simulations are then presented in Sec.~\ref{sec:result}. In Sec.~\ref{sec:resultA}, the dissipation coefficient $\gamma$ is calculated from the simulations with a single heat reservoir. In Sec.~\ref{sec:resultB}, $\gamma_{1,2}$ is estimated from the simulation with two heat reservoirs, from which we can test the additivity of the dissipation coefficients. In Sec.~\ref{sec:resultC}, we present steady-state velocity distributions compared with the Boltzmann distribution. $D_{1,2}$ and $T_{1,2}$ are then estimated. We finally present our full conclusions of the simulations and analysis in Sec.~\ref{sec:conclusion}.

\section{Model}
\label{sec:model}

\begin{figure}
\centering
\includegraphics[width=0.5\linewidth]{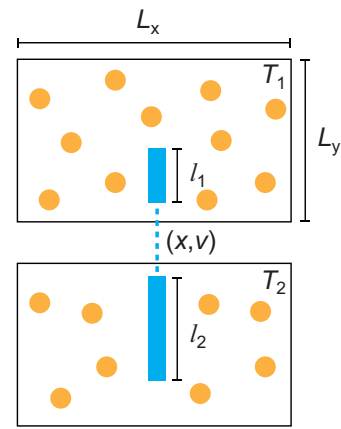}
\caption{(Color online) Schematic of the model. There are two two-dimensional boxes $1$ and $2$ containing $N_1$ and $N_2$ reservoir particles, respectively. The horizontal (vertical) length of the boxes is $L_x$ ($L_y$). The distributions of the reservoir particles in boxes $1$ and $2$ are maintained as in equilibrium at temperatures of $T_1$ and $T_2$, respectively. There is a rod with a length of $l_1$ ($l_2$) in the box $1$ ($2$), and its width is zero. As these two rods are rigidly connected and move only horizontally, their positions $x$ and velocities $v$ are always exactly same. } \label{fig:schematic}
\end{figure}

We adopt the MD simulation model used for the thermal Brownian motor~\cite{Broeck}. Figure~\ref{fig:schematic} shows the schematic of our model, where a system with a single degree of freedom is simultaneously affected by two heat reservoirs. There are two boxes, $1$ and $2$, which contain $N_1$ and $N_2$ reservoir particles, respectively. The boxes are two-dimensional, with horizontal length $L_x$ and vertical length $L_y$. In each of our simulations, we used square boxes, such that $L_x=L_y=L$. The number density of reservoir particles in the box $i$ ($i=1,2$) is $d_i=N_i/L^2$. We also include rods of length $l_1$ and $l_2$ in box $1$ and $2$, respectively, with their widths taken to be zero for simplicity. These two one-dimensional rods are rigidly connected, and can move only horizontally, with no vertical or rotational motion allowed. The motion of the rod set is therefore described by a single degree of freedom, $(x,v)$, where $x$ and $v$ are the position and velocity, respectively, in the horizontal direction only. We refer to this rod set as the rigid stick component, which has a total mass defined as $m$.

Interactions between the reservoir particles are modeled as perfectly elastic hard-disk collisions, with all disks having radius $R$. For simplicity, an elastic collision between a rod and a reservoir particle is assumed to occur when the particle center reaches the rod. We use the Langevin thermostat to maintain the velocity statistics of the reservoir particles for a given temperature $T_i$~\cite{Grest}. That is, their respective motions are determined by the following Langevin equation:
\begin{equation}
\dot{\textbf{\textit{x}}}_\textrm{r}=\textbf{\textit{v}}_\textrm{r},~~m_\textrm{r}\dot{\textbf{\textit{v}}}_\textrm{r} = \textbf{\textit{F}}_\textrm{int} -\gamma_\textrm{r} \textbf{\textit{v}}_\textrm{r} + \boldsymbol{\xi}_{\textrm{r},i}, \label{eq:bath_eq}
\end{equation}
where $\textbf{\textit{x}}_\textrm{r}$ and $\textbf{\textit{v}}_\textrm{r}$ are two-dimensional vectors containing the position and velocity of a reservoir particle, respectively, $m_\textrm{r}$ is mass of a reservoir particle, and $\gamma_\textrm{r}$ is  the dissipation coefficient of the Langevin thermostat. $\boldsymbol{\xi}_{\textrm{r},i}$ is the Gaussian white noise vector of box $i$, satisfying $\langle  \boldsymbol{\xi}_{\textrm{r},i} (t) \boldsymbol{\xi}_{\textrm{r},i}^\intercal (t^\prime) \rangle =2 \gamma_\textrm{r} T_i \delta(t-t') \mathbb{I} $, where $\mathbb{I}$ is the $2\times 2$ identity matrix. $\textbf{\textit{F}}_\textrm{int}$ denotes the interaction forces due to collisions with other reservoir particles, or a rod.
This model has only two reservoirs; however, it is straightforward to extend to a $n$-reservoir model $(n\geq 2)$.

 In all simulations, we implement periodic boundary conditions for each box and
set $m_\textrm{r}=1$, $\gamma_\textrm{r}=1$, $R=0.1$, $m=50$, and $L=30$. We vary other parameters in the range of $0.6 \le T_i \le 1.4$, $0.8 \le l_i \le 1.2$, and $0.2 \le d_i \le 0.3$. Our simulation results shown in the next section provide the numerical estimate of the rigid-rod dissipation coefficient in the range of $0.6 \lesssim \gamma_i  \lesssim 1.2$.

Note that there are four distinct time scales in this model: i) $\tau = m/\gamma_i$ is the relaxation time of the rigid rod, ii) $\tau_\textrm{r} = m_\textrm{r} / \gamma_\textrm{r}$ is the relaxation time of the reservoir particle due to the Langevin thermostat, iii) $\tau_\textrm{s-r} \approx \sqrt{m_\textrm{r}/T_i}(d_i l_i)^{-1}$ is the collision time between the rod and a reservoir particle, and iv) $\tau_\textrm{r-r} = \sqrt{m_\textrm{r}/T_i}(4\sqrt{2}d_i R)^{-1}$ is the collision time between reservoir particles.

For typical values in our simulations ($T_i \sim 1$, $l_i \sim 1$, $d_i \sim 0.25$, and $\gamma_i \sim 1$), the typical time scales are $\tau \sim 50$, $\tau_\textrm{r} \sim 1$, $\tau_\textrm{s-r} \sim 4$, and $\tau_\textrm{r-r} \sim 7$.
First, we point out that, with $\tau_\textrm{s-r} \approx 4 \tau_\textrm{r}$, the memory of the previous collision between the rod and a reservoir particle should be considerably weakened when they collide with each other again at the next collision. Second,
with $\tau/\tau_\textrm{s-r} \sim 10$, these time scales are well separated, but maybe higher collision statistics are necessary to
match the Langevin equation~\eqref{eq:Lan_eq1} perfectly without any transient period. We will see multiple relaxation modes
at early times from simulations in the next section.
Finally, we note that $\tau_\textrm{r-r}$ is irrelevant due to the Langevin thermostat with much shorter relaxation time
($\tau_\textrm{r} \ll \tau_\textrm{r-r}$).



\section{Simulations and Results}
\label{sec:result}

To verify the additivity of multiple reservoirs in the Langevin equation, we examine two theoretically additive properties. These are the dissipation coefficients, such that $\gamma_{1,2} = \gamma_1 + \gamma_2$, and the noise strengths, such that $D_{1,2}=D_1 + D_2$.

First, we calculate the dissipation coefficient $\gamma_i$ from Eq.~\eqref{eq:Lan_eq1} using the MD simulations, in which the rigid stick is connected to a single heat reservoir $i$ ($i=1,2$). Analysis results are presented in Sec.~\ref{sec:resultA}, where we measure $\gamma_i$ from finite-time relaxation dynamics. Following this, we repeat the procedure for the BM dynamics where the rigid stick is in contact with heat reservoirs $1$ and $2$ simultaneously. The dissipation coefficient $\gamma_{1,2}$ in Eq.~\eqref{eq:Lan_eq2} is measured and compared with the individual case to verify the additive relationship $\gamma_1+\gamma_2 = \gamma_{1,2}$. This analysis is presented in Sec.~\ref{sec:resultB}. We also examine other relaxation modes corresponding to short-time dynamics for possible corrections to Eq.~\eqref{eq:Lan_eq2}.

Second, the additivity of noise strength can be measured, and possibly confirmed, using the following procedure. We obtain the steady-state probability distribution function for BM and compare it with the expected Boltzmann (Gaussian) distribution. Then, we estimate $T_{1,2}$ from the velocity distribution, which yields $D_{1,2}$ via the relation of $D_{1,2}=\gamma_{1,2} T_{1,2}$.

\subsection{Dissipation coefficient with a single reservoir}
\label{sec:resultA}

\begin{figure}
\centering
\includegraphics[width=0.9\linewidth]{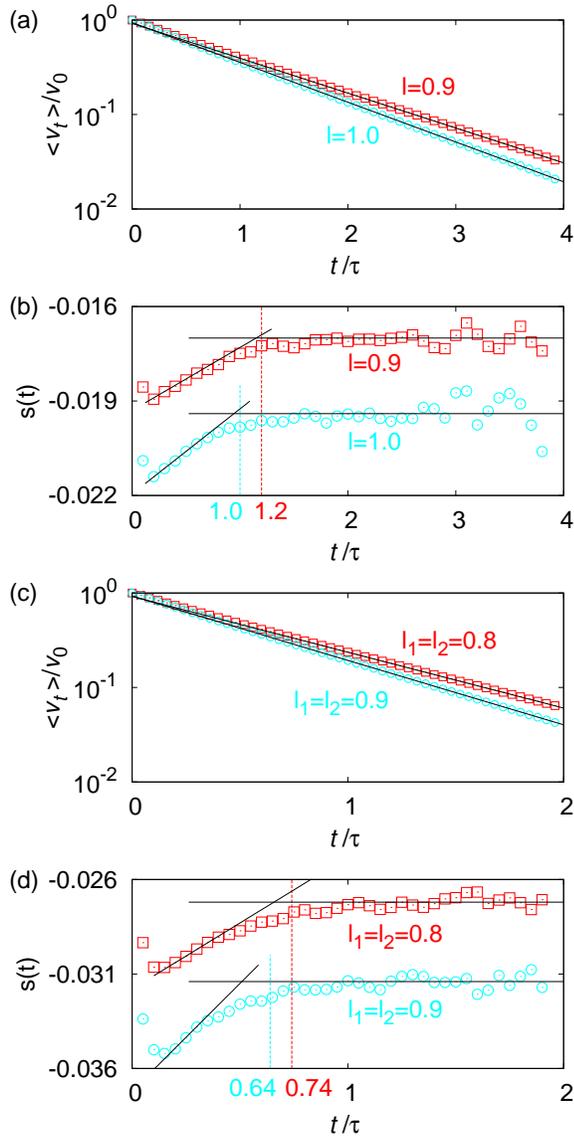}
\caption{(Color online) Estimates for the dissipation coefficients. (a) Single reservoir at $T=1.0$. Semi-log plots of $\langle v_t \rangle/v_0$ versus $t/\tau$ for $l=0.9$ and $1.0$ with $d=0.25$ with the rigid-rod relaxation time $\tau$. Slopes of the solid lines are determined from the long-time values of Fig.~\ref{fig:fitting}(b). (b) Plots of successive slope $s(t)$ of data in (a) against $t/\tau$.
The horizontal lines  denote  the long-time average slope $s_\textrm{s}$ and the vertical lines indicate
the saturation time $\tau_\textrm{s}$.
(c) Two reservoirs at $T_1=1.0$ and $T_2=0.6$, respectively. Semi-log plots of $\langle v_t \rangle/v_0$ versus $t/\tau$ for $l_1=l_2=0.8$ and $0.9$ with $d_1=d_2=0.25$. Slopes of the solid lines are determined from the long-time values of Fig.~\ref{fig:fitting}(d).
(d) Plots of successive slope $s(t)$ of data in (c) against $t/\tau$.
} \label{fig:fitting}
\end{figure}

\begin{figure}
\centering
\includegraphics[width=0.99\linewidth]{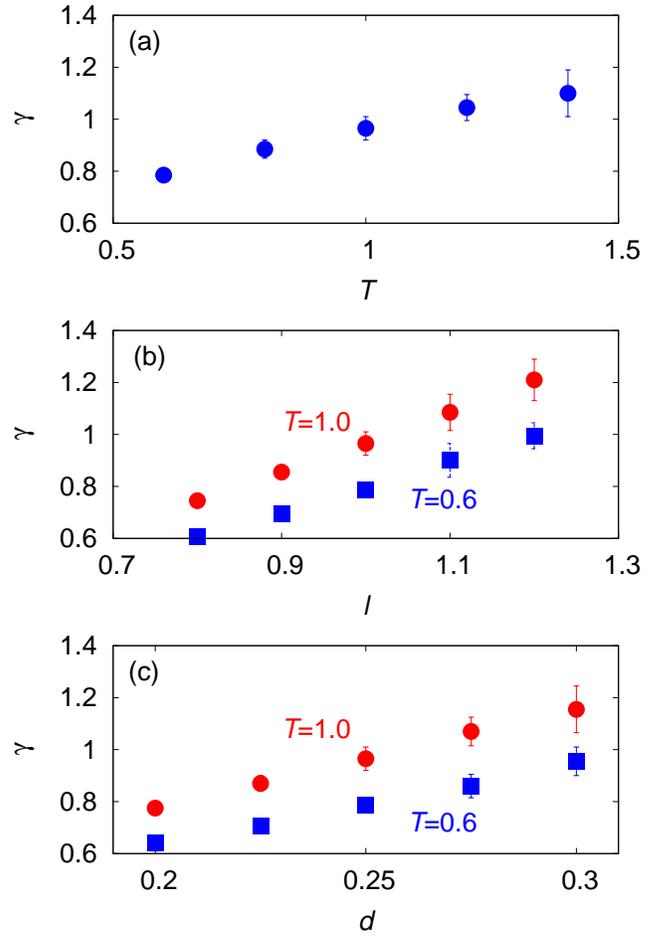}
\caption{(Color online) $T$, $l$, and $d$ dependence of $\gamma$.
(a) $T$ dependence of $\gamma$ with $l=1.0$ and $d=0.25$. (b) $l$ dependence of $\gamma$ for $T=1.0$ and $0.6$ with $d=0.25$. (c) $d$ dependence of $\gamma$ for $T=1.0$ and $0.6$ with $l=1.0$. } \label{fig:onebath}
\end{figure}

Here, we consider the case of a single heat reservoir. This simulation is possible in our setup when $N_2=0$, meaning that the rigid stick is in contact with reservoir $1$ only. The set of control parameters are $T_1$, $l_1$, and $d_1$; for the remainder of this section, for convenience, we remove the subscript notation and denote them as $T$, $l$, and $d$, respectively. Initial velocity is set as $v_0 = 1$ for the following simulations unless otherwise noted.

The rigid stick motion is supposed to be described by the Langevin equation~\eqref{eq:Lan_eq1} with zero external force, i.e., $f(x)=0$. By taking averages of both sides of the Langevin equation, we obtain $m \langle \dot{v}_t \rangle = -\gamma \langle v_t \rangle$, where we have omitted the dummy subscript $i$. Then, the velocity relaxation dynamics are given by
\begin{equation}
\langle v_t \rangle = \langle v_0 \rangle e^{-\frac{\gamma}{m}t}, \label{eq:time_evolution}
\end{equation}
where $\langle v_0 \rangle =1$ and $m=50$.

In simulations, we measure the time dependence of $\langle v_t \rangle $  up to $t=200$ and estimate the dissipation coefficient $\gamma$ numerically.
Figure~\ref{fig:fitting} shows its estimate procedure in details. Figure~\ref{fig:fitting}(a) displays semi-log plots of $\langle v_t \rangle$ against $t$ for two different values of $l=0.9$ and $l=1.0$ with $T=1.0$ and $d=0.25$. We approximate the ensemble average $\langle v_t\rangle$ by averaging over $10^5$ simulation runs.
Overall data seem to be well fitted by a linear regression model as expected from Eq.~\eqref{eq:time_evolution}. However, there is a slight deviation at early times. To carefully analyze this initial transient behavior, we calculate the successive slopes, defined as $s(t)\equiv (\ln\langle v_{t+\delta} \rangle -\ln\langle v_{t} \rangle )/\delta$, which are presented in Fig.~\ref{fig:fitting}(b) with $\delta =5$.
As can be seen in this figure, the slope changes at early times but saturates in the long-time limit. We estimate the long-time slope
$s_\textrm{s}$ by averaging over slope data in the saturated regime ($t>\tau_\textrm{s}$) and determine the dissipation coefficient value by $\gamma=-ms_\textrm{s}$.
In all figures, the time axis is shown in the unit of the typical rigid-rod relaxation time $\tau=m/\gamma \approx 50$, where
one can see that the saturation regime starts around $\tau_\textrm{s} /\tau\sim 1$.
The horizontal lines in Fig.~\ref{fig:fitting}(b) denote  $s_\textrm{s}$ and the vertical lines indicate  $\tau_\textrm{s}$.

We note that there are multiple relaxation modes in this dynamics for large but finite $\tau/\tau_\textrm{s-r}$~\cite{Meurs}. The dominant (first) mode is described by the Langevin equations~\eqref{eq:Lan_eq1} or \eqref{eq:Lan_eq2} with the relaxation time $m/\gamma$. The relaxation time of the second mode is $m/(2\gamma)$ and higher modes have shorter relaxation times. Thus, it is expected that all the modes except the dominant one are almost invisible for $t \gtrsim \tau \sim \tau_\textrm{s}$.  For $l=0.9$ and $1.0$ cases, we find $\tau_\textrm{s}/\tau \simeq 1.2$ and $1.0$, respectively, as seen in Fig.~\ref{fig:fitting}(b).

We study  the temperature dependence of the dissipation coefficient. In this simulation, we vary $T$ with fixed $l=1.0$ and $d=0.25$.
From the saturated slopes, we estimate the values of dissipation coefficients for various values of $T$, which are presented in Fig.~\ref{fig:onebath}(a). We see that the dissipation coefficient increases as the temperature increases. Similar tendency can be also found from Sutherland's formula for ideal gases~\cite{Smits}.

We then proceed to study the rod-length dependence of the dissipation coefficient. In this simulation, we fix $d=0.25$ and $T=1.0$ or $0.6$ while varying $l$. Again, we obtain the saturated slops for all values of $l$. The results can be seen in Fig.~\ref{fig:onebath}(b). We find that the relaxation dynamics become more dissipative for longer lengths of rod. This is easily understood by the Stokes' law; the dissipation coefficient is proportional to the radius of a Brownian particle~\cite{Stokes}.

Finally, we investigate the dependence on the density of reservoir particles, $d$, which is shown in Fig.~\ref{fig:onebath}(c). In this simulation, we set $l=1.0$ and $T=1.0$ or $T=0.6$. As seen in the figure, the dissipation coefficient increases with the density. This is expected, as a more crowded environment of reservoir particles increases the total resistance to the rod motion.

In summary, the dissipation coefficient in the presence of a single reservoir $i$ is a function of $T_i$, $l_i$, and $d_i$, i.e., $\gamma_i = \gamma_i (T_i, l_i, d_i)$. For use in following sections, we define the notation $\gamma_i (z_i)$, where $z=T ,l, d$. This allows us to denote $z_i$ as the only varying parameter, where two of its parameters are fixed at given values.

\begin{figure*}
\centering
\includegraphics[width=0.96\textwidth]{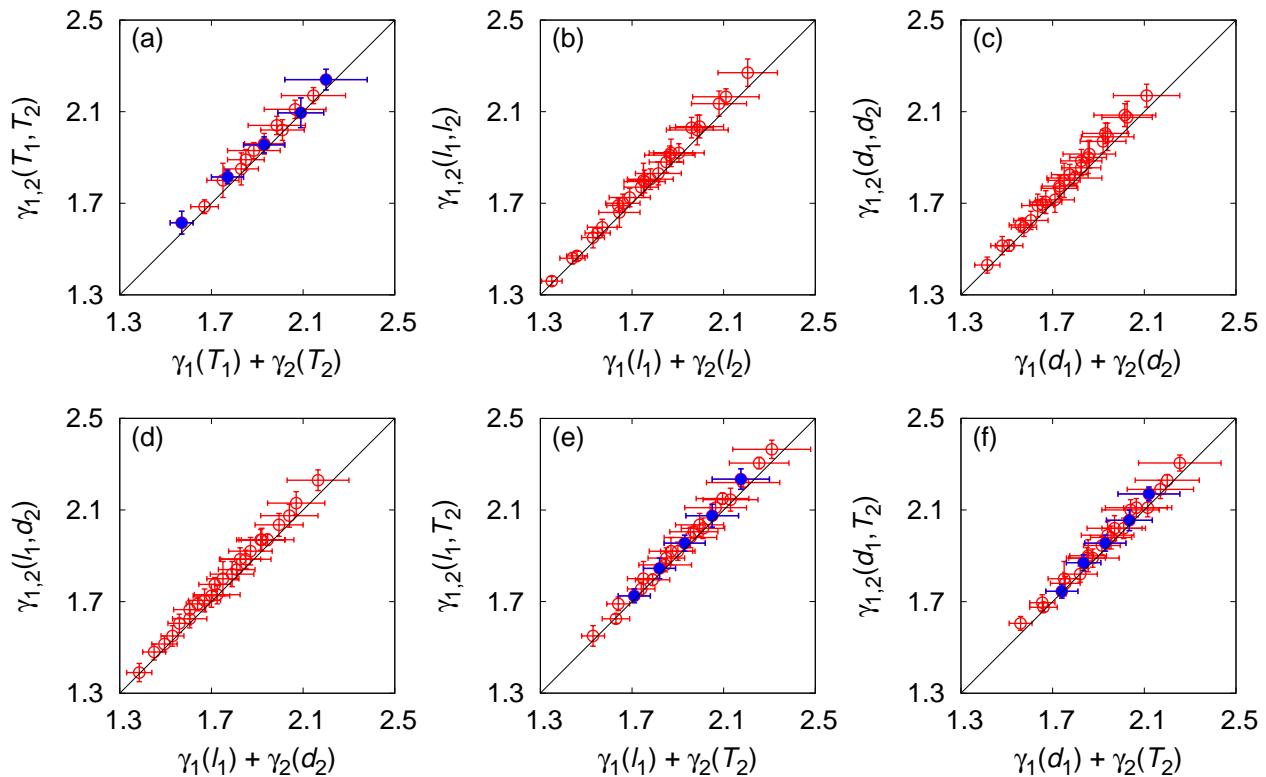}
\caption{(Color online) Additivity of dissipation coefficients. (a) Plot of $\gamma_{1,2}$ versus $\gamma_1 +\gamma_2$ when $T_1$ and $T_2$ are varied. Statistical errors are denoted by small vertical and horizontal ticks, and the solid line denotes the `$y=x$' line. (b)  Plot of $\gamma_{1,2}$ versus $\gamma_1 +\gamma_2$ when $l_1$ and $l_2$ are varied. (c)  Plot of $\gamma_{1,2}$ versus $\gamma_1 +\gamma_2$ when $d_1$ and $d_2$ are varied. (d) Plot of $\gamma_{1,2}$ versus $\gamma_1 +\gamma_2$ when $l_1$ and $d_2$ are varied. (e) Plot of $\gamma_{1,2}$ versus $\gamma_1 +\gamma_2$ when $l_1$ and $T_2$ are varied. (f) Plot of $\gamma_{1,2}$ versus $\gamma_1 +\gamma_2$ when $d_1$ and $T_2$ are varied. In all the plots, filled circles represent the data for $T_1=T_2$. } \label{fig:twobath}
\end{figure*}

\subsection{Dissipation coefficient with two reservoirs}
\label{sec:resultB}

We now consider the case of two reservoirs, where $N_1, N_2 \neq 0$, and the rigid stick is in contact with both heat reservoirs $1$ and $2$ simultaneously. If the rigid stick motion in this situation can, indeed, be described by Eq.~\eqref{eq:Lan_eq2} after the initial transient regime, we would expect the average velocity to show the following relaxation behavior:
\begin{equation}
\langle v_t \rangle = \langle v_0 \rangle e^{-\frac{\gamma_{1,2}}{m}t}. \label{eq:time_evolution_2}
\end{equation}
To determine whether $\langle v_t \rangle$ corresponds to Eq.~\eqref{eq:time_evolution_2}, we performed extensive MD simulations for many different pairs of $(T_1, l_1, d_1, T_2, l_2, d_2)$. Figure~\ref{fig:fitting}(c) shows semi-log plots of $\langle v_t \rangle$ against $t$ for two values of $l_1=l_2=0.8,~0.9$ at fixed $T_1=1.0$, $T_2=0.6$, and $d_1=d_2=0.25$. Following the same procedure as done for obtaining Fig.~\ref{fig:fitting}(b), we obtain the saturated slopes from Fig.~\ref{fig:fitting}(d). The solid lines in Fig.~\ref{fig:fitting}(c) are drawn by using these saturated slopes.
After the transient regime, we see that all data are well fitted by the straight lines. The saturation time $\tau_\textrm{s}/\tau$ for $l_1=l_2=0.8$ and $0.9$ are estimated as $0.74$ and $0.64$, respectively  in Fig.~\ref{fig:fitting}(d). We also obtain data for many other pairs of $(T_1, l_1, d_1, T_2, l_2, d_2)$, which again correspond well to fitted linear regression. From the saturated slopes, we determine the values of $\gamma_{1,2}$ as a function of parameters $(T_1, l_1, d_1, T_2, l_2, d_2)$. Using our compact notation, we define $\gamma_{1,2} (z_1,z_2)$, where $z= l, T, d$, denoting that only $z_1$ and $z_2$ are varying parameters, whereas the others are fixed at given values.

\begin{figure}
\centering
\includegraphics[width=0.49\textwidth]{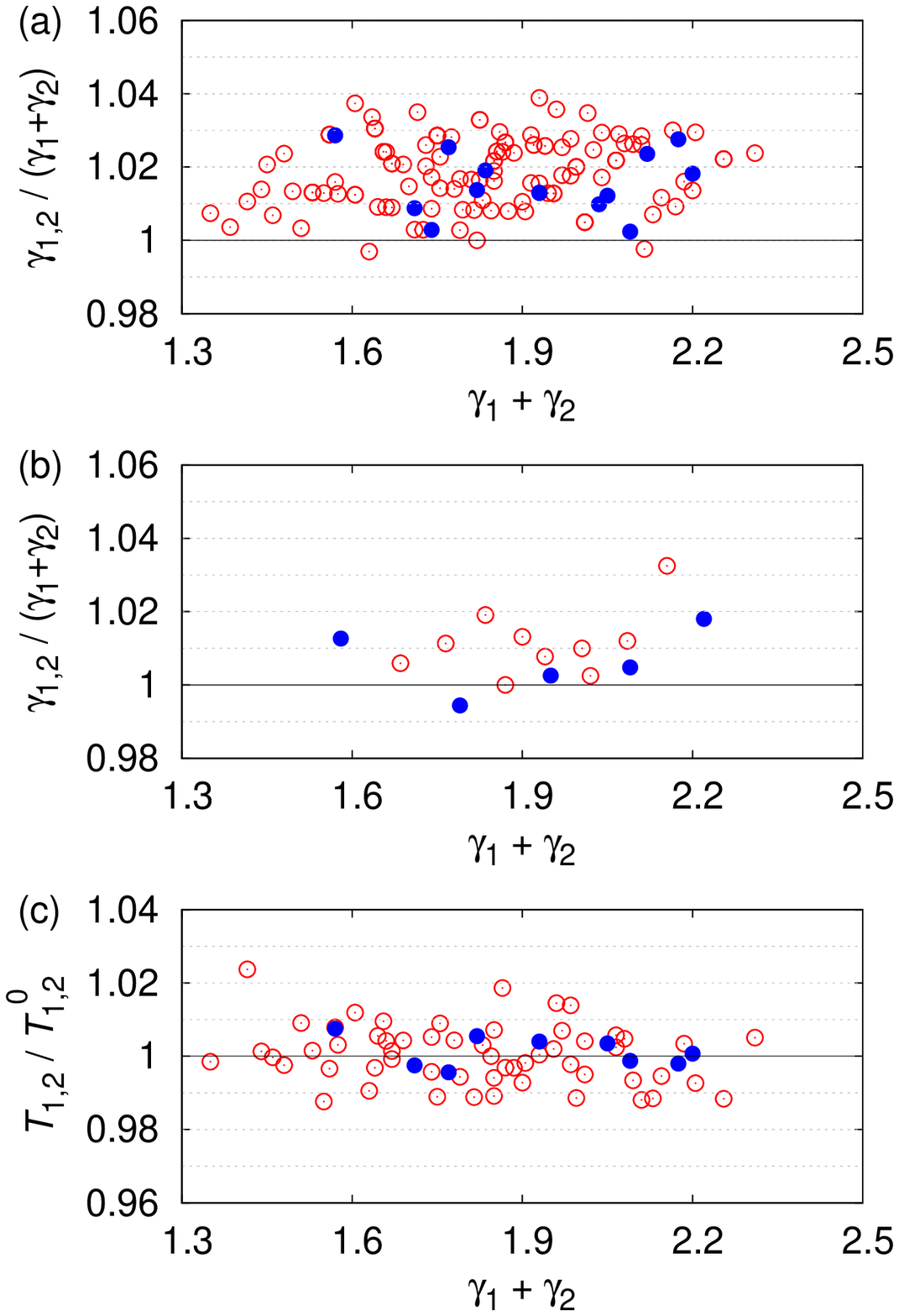}
\caption{(Color online) Statistical deviation from the simple additivity. (a) Dissipation coefficients with $v_0=1$. (b) Dissipation coefficients with the steady-state initial condition.
(c) Effective temperatures. Filled circles represent the data for $T_1=T_2$.  } \label{fig:ratio}
\end{figure}

Now, we analyze the additivity of dissipation coefficients; that is, we examine whether $\gamma_{1,2} (z_1, z_2) = \gamma_1 (z_1)+\gamma_2 (z_2)$. First, we fix $l_1=l_2=1.0$ and $d_1=d_2=0.25$ and vary $T_1$ and $T_2$. Figure~\ref{fig:twobath}(a) shows the plots of $\gamma_{1,2} (T_1,T_2)$ against $\gamma_1 (T_1)+\gamma_2(T_2)$ for various pairs of $(T_1,T_2)$, with $T_1$ and $T_2$ taking the values $0.6, 0.8, 1.0, 1.2$, and $1.4$. We see that the data trend is again well fitted by a straight line.
There seems a slight deviation from the exact additivity in all data, which we write as
\begin{equation}
\frac{\gamma_{1,2}}{\gamma_1 +\gamma_2}=1+\epsilon, \label{eq:epsilon}
\end{equation}
where $\epsilon$ is positive and its magnitude is about $2\%$.
However, the deviation magnitude is smaller than statistical errors about $\sim 4\%$, so it is difficult to figure out its origin whether it comes from an measurement artifact or has a meaningful physical mechanism.
One thing we want to mention is that this slight deviation cannot be attributed to non-equilibrium behavior for $T_1 \neq T_2$, where an energy current between two reservoirs through the rigid stick is expected. We find that, even for the equilibrium situation, $T_1=T_2$, (perfectly identical reservoirs with the same $d$ and $l$), $\epsilon$ remains, with a similar magnitude as in the $T_1 \neq T_2$ case. These are represented by five filled dots in Fig.~\ref{fig:twobath}(a).

\begin{figure*}
\centering
\includegraphics[width=0.99\textwidth]{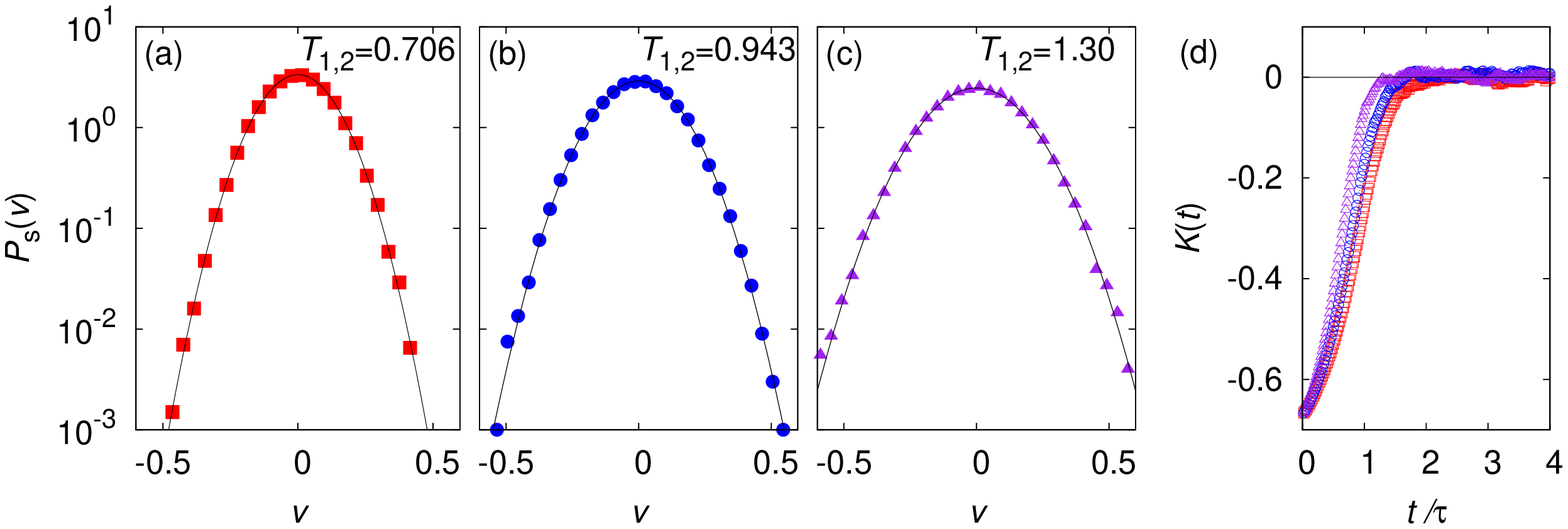}
\caption{(Color online) Steady state velocity distribution. Filled dots represent numerical results for (a) $(T_1,T_2)=(0.8, 0.6)$, (b) $(T_1,T_2)=(1.2,0.6)$, and (c) $(T_1,T_2)=(1.4,1.2)$ with $l_1=l_2=1$ and $d_1=d_2=0.25$. Solid curve denotes the Boltzmann distribution Eq.~\eqref{eq:Boltzmann} with (a) $T_{1,2}=0.706$, (b) $T_{1,2}=0.943$. (c) $T_{1,2}=1.30$. (d) Scaled kurtosis curves as a function time for (a), (b), and (c) cases.  } \label{fig:steady}
\end{figure*}

We also check the additive property of dissipation coefficients with various parameter values. Figure~\ref{fig:twobath}(b) shows the plots for fixed values of $d_1=d_2=0.25$, $T_1=1.0$, and $T_2=0.6$ and various pairs of $(l_1,l_2)$ with $l_1, l_2= 0.8, 0.9, 1.0, 1.1$, and $1.2$. For Fig.~\ref{fig:twobath}(c), we fix $l_1=l_2=1.0$, $T_1=1.0$, and $T_2=0.6$ for pairs of $(d_1,d_2)$ with $d_1, d_2=0.2, 0.225, 0.25, 0.275$, and $0.3$. In Figs.~\ref{fig:twobath}(d), \ref{fig:twobath}(e), and \ref{fig:twobath}(f), more general situations are considered, where pairs of $(l_1, d_2)$, $(l_1, T_2)$, and $(d_1,T_2)$ are varied, respectively. In all cases, the additivity behavior is clearly satisfied with the similar small deviation. The deviation magnitude is better seen in Figure~\ref{fig:ratio}(a), showing plots of the same data  in Fig.~\ref{fig:twobath} in terms of the ratio of $\gamma_{1,2}$ and $\gamma_1 +\gamma_2$. The average value of the ratio is $1.02$ ($2\%$ deviation from the additivity), which is smaller than  statistical errors ($4\% \sim 7\%$).

Note that all simulations so far start from the initial condition, $v_0=1$. This initial condition sets a very high initial energy ($=25$) of the rigid stick compared with the thermal energy ($\sim 1$). One might suspect that this unusual initial condition could affect the relaxation dynamics. To check this, we perform simulations with the steady-state initial condition, where we expect from the Langevin equations as
\begin{equation}
\langle v_t v_0 \rangle = \langle v_0^2 \rangle e^{-\frac{\gamma}{m} t}.
\end{equation}
We choose the same parameters used in Fig.~\ref{fig:twobath}(a) and obtain the average values of the correlation function over $5\times 10^5$ samples. Figure~\ref{fig:ratio}(b) shows the simulation result, which seems similar to the previous case with the $v_0=1$ initial condition.

\subsection{ Steady-state distribution and effective temperature with two reservoirs}
\label{sec:resultC}

Here, we check the steady-state distribution with two reservoirs. The velocity distribution $P(v,t)$ of Eq.~\eqref{eq:Lan_eq2} can be calculated from the corresponding Fokker-Plank equation which is given by
\begin{equation}
\frac{\partial}{\partial t}P(v,t) = \frac{\partial}{\partial v} \left( \frac{\gamma_{1,2}}{m}v +\frac{D_{1,2}}{m^2}\frac{\partial}{\partial v}  \right) P(v,t). \label{eq:Fokker}
\end{equation}
The steady state distribution of Eq.~\eqref{eq:Fokker} is the following Boltzmann distribution:
\begin{equation}
P_\textrm{s} (v) = \sqrt{\frac{m}{2\pi T_{1,2}}} e^{-\frac{m v^2}{2 T_{1,2}}}, \label{eq:Boltzmann}
\end{equation}
where $T_{1,2} = D_{1,2}/\gamma_{1,2}$. Note that the effective temperature $T_{1,2}$ is neither $T_1$ nor $T_2$, so the rigid stick is not in an equilibrium state; indeed, it is in a non-equilibrium steady state, even though the distribution is a Boltzmann distribution. As a result, there is a finite heat current~\cite{Parrondo, Visco} and positive entropy production~\cite{Broeck1,Murashita} in BM.

If the additivity of the noise magnitudes ($D_{1,2}=D_1+D_2$) were assumed to be valid with the additivity of dissipation coefficients ($\gamma_{1,2}=\gamma_1 +\gamma_2$), $T_{1,2}$ becomes
\begin{equation}
T_{1,2}=\frac{\gamma_1 T_1+ \gamma_2 T_2}{\gamma_1 + \gamma_2} \equiv T_{1,2}^0
. \label{eq:T12}
\end{equation}
However, as there is a small correction in the additivity of dissipation coefficients as in Eq.~\eqref{eq:epsilon}, it is not clear that $T_{1,2}$ is equal to $T_{1,2}^0$ without any correction.

We perform MD simulations to check the validity of Eqs.~\eqref{eq:Boltzmann} and \eqref{eq:T12} from the steady-state distributions. We use $57$ parameter sets of $(T_1, l_1, d_1, T_2, l_2, d_2)$ for $T_1 \neq T_2$ cases and $13$ parameter sets for $T_1 = T_2$ cases. As the relaxation time scale $m/\gamma_{1,2} \sim 30$, we gather $5\times 10^4$ sets of velocity data, starting from $t=5000$, to obtain the steady-state distribution.

Figure~\ref{fig:steady}(a), (b), and (c) show the examples of the simulated (dots) and expected (solid curves) distributions of $v$ for three parameter sets with $T_1 \neq T_2$. In these sets, we fix $l_1=l_2=1$ and $d_1=d_2=0.25$ and take three different pairs of $(T_1,T_2)$, such as $(0.8,0.6),$ $(1.2,0.6)$, and $(1.4,1.2)$, which set $(\gamma_1,\gamma_2)$ as $(0.885, 0.785)$, $(1.05, 0.785)$, and $(1.10, 1.05)$, respectively. For these pairs, Eq.~\eqref{eq:T12} predicts $T_{1,2}^0=0.706$, $0.943$, and $1.30$, respectively. The simulated distributions are reasonably well fitted by the expected Boltzmann curves, implying that $T_{1,2}\simeq T_{1,2}^0$. We check non-Gaussianity quantitatively by measuring the scaled kurtosis, $K(t)=\langle v_t^4 \rangle/(3\langle v_t^2\rangle^2)-1$, as a function of time, which are presented in Fig.~\ref{fig:steady}(d). Note that $K(t)$ approaches zero as the distribution goes toward the Gaussian.  After $t/\tau \sim 2$, all the scaled kurtosis curves converge to zero with statistical errors less than $0.006$  for all cases, which confirm the Gaussian steady-state
distribution of Eq.~\eqref{eq:Boltzmann}.

We also measure the velocity dispersion $\langle v^2\rangle_\textrm{s}$ for each steady state distribution and estimate $T_{1,2}$ quantitatively using the relation $T_{1,2} \equiv m\langle v^2\rangle_\textrm{s}$ given by Eq.~\eqref{eq:Boltzmann}. Figure~\ref{fig:ratio}(c) shows the ratio $T_{1,2}/T_{1,2}^0$ for all $70$ parameter sets. In contrast to the case of dissipation coefficients in Fig.~\ref{fig:ratio}(a), there seems no systematic deviation such that the average value of the ratio is $1.005$
 ($0.5\%$ deviation from the additivity), which is much smaller than statistical errors  ($\sim 3\%$).
 Thus, we conclude that the steady-state is almost perfectly described by Eq.~\eqref{eq:Lan_eq2}, supporting the claim by H\"{a}nggi~\cite{Hanggi1}.


\section{Conclusions}
\label{sec:conclusion}

In this work, we investigate the additivity of two heat reservoirs with arbitrary temperatures by extensive MD simulations. We first estimate dissipation coefficients from the relaxation dynamics and check the additivity of dissipation coefficients, $\gamma_{1,2}=\gamma_1+\gamma_2$. We find that the additivity is satisfied very well in general only with a small deviation less than statistical errors. The origin of this small deviation is unclear whether it is resulted from a measurement artifact or a certain physical mechanism.
Nevertheless, as its magnitude is smaller than the numerical accuracy, we conclude that the additivity is at least `statistically' valid.
In addition, we find that the initial-condition dependence of the relaxation dynamics is not substantial to be considered. Therefore,
 concern about the initial-condition dependence raised by H\"{a}nggi~\cite{Hanggi1} and Parrondo and Espa\~{n}ol~\cite{Parrondo} 
 in non-equilibrium situations may be regarded not significant in general. Finally, we report that the steady-state distribution satisfies the additivity almost perfectly, as expected.


\begin{acknowledgments}
	This research was supported by the NRF grant No.~2011-35B-C00014 (JSL) and 2017R1D1A1B06035497 (HP).
\end{acknowledgments}

\appendix

\vfil\eject

\end{document}